# Revisiting Exception Handling Practices with Exception Flow Analysis

Guilherme B. de Pádua
Department of Computer Science and Software Engineering
Concordia University - Montreal, QC, Canada
Email: g_bicalh@encs.concordia.ca

Weiyi Shang
Department of Computer Science and Software Engineering
Concordia University - Montreal, QC, Canada
Email: shang@encs.concordia.ca

*Abstract*—Modern programming languages, such as Java and C#, typically provide features that handle exceptions. These features separate error-handling code from regular source code and aim to assist in the practice of software comprehension and maintenance. Having acknowledged the advantages of exception handling features, their misuse can still cause reliability degradation or even catastrophic software failures. Prior studies on exception handling aim to understand the practices of exception handling in its different components, such as the origin of the exceptions and the handling code of the exceptions. Yet, the observed findings were scattered and diverse. In this paper, to complement prior research findings on exception handling, we study its features by enriching the knowledge of handling code with a flow analysis of exceptions. Our case study is conducted with over 10K exception handling blocks, and over 77K related exception flows from 16 open-source Java and C# (.NET) libraries and applications. Our case study results show that each try block has up to 12 possible potentially recoverable yet propagated exceptions. More importantly, 22% of the distinct possible exceptions can be traced back to multiple methods (average of 1.39 and max of 34). Such results highlight the additional challenge of composing quality exception handling code. To make it worse, we confirm that there is a lack of documentation of the possible exceptions and their sources. However, such critical information can be identified by exception flow analysis on well-documented API calls (e.g., JRE and .NET documentation). Finally, we observe different strategies in exception handling code between Java and C#. Our findings highlight the opportunities of leveraging automated software analysis to assist in exception handling practices and signify the need of more further in-depth studies on exception handling practice.

## I. INTRODUCTION

Modern programming languages, such as Java and C#, typically provide exception handling features, such as throw statements and try-catch-finally blocks. These features separate error-handling code from regular source code and are leveraged widely in practice to support software comprehension and maintenance [1], [2].

Having acknowledged the advantages of exception handling features, their misuse can still cause catastrophic software failures, such as application crashes [3], or reliability degradation, such as information leakage [4], [5]. A large portion of systems has suffered from system crashes that were due to exceptions [6]. Additionally, the importance of exception handling source code has been illustrated in prior research and surveys [7], [8].

Prior studies on exception handling aim to understand the practices of exception handling in its different components: exception sources and handling code. Yet, the observed findings were scattered and diverse. Recent empirical studies on exception handling practices have advocated the suboptimal use of exception handling features in open source software [9], [10], [11], [12]. Moreover, in our previous research, we observe the prevalence of exception handling anti-patterns. These research findings imply the lack of a thorough understanding of the practice of exception handling [13].

Therefore, in this paper, we re-visit exception handling practices by conducting an in-depth study on 16 open-source Java and C# libraries and applications. To understand and analyze the state-of-the-practice of exception handling in these projects, we perform source code analysis to track the flow of exceptions from the source of exceptions, through method invocations, to the attempting blocks of exceptions (try block) and the exception handling block (catch block). With such flow analysis, we extract information about exception handling practice in over 10K exception handling blocks, and over 77K related exception flows from the studied subject systems.

Our case study focuses on four aspects of the exception handling practices: 1) the quantity of exceptions, 2) the diversity of exceptions, 3) the sources of exceptions and 4) the exception handling strategies and actions. Table I summarizes our findings and their corresponding implications. Such results confirm the challenge of composing quality exception handling code. For example, we find a considerable amount of potentially recoverable yet propagated exceptions. However, more importantly, we highlight the opportunities of leveraging our automated source code analysis to complement the information that is valuable for developers when handling exceptions. More in-depth analyses are needed to ensure and improve the quality and usefulness of exception handling in practice.

The rest of the paper is organized as follows: Section II discusses the related prior research of this work. Section III presents the methodology of the exception flow analysis through an illustrative example and our case study setup. Section IV to VII presents the results of our case study. Section VIII discusses the threats to the validity of our findings. Finally, Section IX concludes the paper and discusses potential future research directions based on our research results.

TABLE I
OUR FINDINGS AND IMPLICATIONS ON EXCEPTION HANDLING PRACTICES.

| Quantity of Exceptions (Section IV) | Implications |
| --- | --- |
| (1) There often exist multiple possible exceptions in each try block, and, out of those, many are propagated. | Current state-of-the-practice may not provide information to developers about all possible exceptions. Automated techniques may help developers be aware of all possible exceptions to make exception handling decisions. |
| (2) There exists a considerable amount of potentially recoverable exceptions that are propagated, even though they are recommended to be handled by Java and C#. | Exception flow analysis can provide automated tooling support to alert developers about not handling potentially recoverable exceptions. |
| **Diversity of Exceptions (Section V)** | **Implications** |
| (3) With a significant amount of exceptions existing in each project, many possible exception types appear in only one try block. | Developers may not need to be aware of all exception types in a project by receiving automated suggestions of the exceptions that he/she needs to understand. |
| **Sources of Exceptions (Section VI)** | **Implications** |
| (4) Over 22% of the exceptions are traced from different methods. | Automated tools are needed to help developers understand the source of the exception if it is traced back to different methods. |
| (5) The libraries used by the systems can provide documentation to most of the possible exceptions. | Developers should leverage automated analyses to understand possible exceptions. |
| **Exception Handling Strategies and Actions (Section VII)** | **Implications** |
| (6) Only a small portion of the exceptions are handled with the *Specific* strategy. | Developers should be guided to prioritize on handling exceptions with the *Specific* strategy, since developers cannot optimize the handling of the exception without knowing its exact type information. |
| (7) Java and C# have differences in leveraging various actions when handling exceptions. | More in-depth analysis and user studies are needed to further understand the rationale of differences of Java and C# exception handling practices. |
| (8) Actions that are taken when handling exceptions with specific or subsumption manners are not statistically significantly different. | Research and tooling support are needed to guide how to handle exceptions, especially with the specific strategy. |
| (9) With statistical significance, all top 10 Java and 2 out of top 10 C# exceptions have at least one action that is taken differently from the rest of the exceptions. | Developers may consider leveraging automated suggestion of exception handling actions. |

## II. RELATED WORK

### A. Empirical studies on exception handling practices

Prior research studied exception handling based on source code and issue trackers. Cabral and Marques [14] studied exception handling practices from 32 projects in both Java and .Net without considering the flow of exceptions. Prior work by Jo *et al.* [15] focuses on uncaught exceptions of Java Checked exceptions. They proposed an inter-procedural analysis based on set-based framework without using declared exceptions.

Coelho *et al.* [16] assessed exception handling strategy with exception flows from Aspect-oriented systems and object-oriented systems. They evaluated the number of *uncaught* exceptions, exceptions caught by *subsumption*, and exceptions caught with *specialized* handlers.

Sena *et al.* [17] investigated sampled exception flows from 656 Java libraries for flow characteristics, handler actions, and handler strategies. We extend their work by looking into a higher number of flows per system (e.g. in Apache ANT we identified 930 catch blocks, compared to 2), by considering applications besides libraries and including C# .NET systems.

Some studies reveal that developers consider exception handling hard to learn and to use and tend to avoid it or misuse it [9], [10], [11]. Bonifacio *et al.* [12] also surveyed C++ developers encountering revelations of educational issues.

It also has been noted that there is a lack of documentation of exceptions. Kechagia and Spinellis [18] found that 69% of the methods had undocumented exceptions and 19% of crashes could have been caused by insufficient documentation. Sena *et al.* [17]'s findings confirm that API runtime exceptions are poorly documented. Cabral and Marques [19] identify that infrastructure (20%) and libraries (15%) have better exception handling documentation when compared to applications (2%).

Significant research aimed to indicate exception handling problems and their impacts. Sinha *et al.* [20] leveraged exception flow analyses to study the existence of 11 anti-patterns in four Java systems. Other research [2], [8], [7] classified exception-handling related bugs by mining software issue tracking. In our previous research, we observe the prevalence of exception handling anti-patterns [13].

Cacho *et al.* [21], [22] studied the evolution of the behavior of exception handling in Java and C# source code changes. Their results highlight the impact of the programming language design differences in the maintenance and robustness of exception handling mechanisms.

Our study revisits and combines different aspects of the studies mentioned above. Moreover, we present new findings (see Table I) that are not yet highlighted in prior research.

## B. Improving exception handling practices

Robillard and Murphy [23] created a tool to analyze exception flows in Java programs, including a graphical user interface. Similarly, Garcia and Cacho [24] proposed a different approach for .NET related languages. Garcia and Cacho's tool supports visualization of metrics over the application history.

To support the software development lifecycle, Sinha *et al.* [20] provided automated support for development, maintenance and testing requirements related to exception handling.

To improve how developers would deal with exception handling complexity, Kechagia *et al.* [25] discuss and propose improvements in the design of exception handling mechanisms. Zhang and Krintz [26] propose an as-if-serial exception handling mechanism for parallel programming.

The burden of writing exception handling code has been pointed out by Cabral and Marques [27]. They showed that a system with an automated set of recovery actions is capable of achieving better error resilience than a traditional system.

Barbosa *et al.* developed strategies with heuristics for recommending exception handling code as a semi-automated approach. Zhu *et al.* [28] proposed an approach that suggests logging decisions for exception handling.

By conducting an in-depth study on 16 open source projects, our findings illustrate the opportunities of leveraging various analysis to combine information from different sources to understand and assist in exception handling flows and practices. Our results are valuable to complement and assist in improving existing exception handling techniques.

## III. METHODOLOGY

In this section, we present the methodology of our study. Aiding the explanation of our methodology, we first consider an illustrative example. Second, we introduce our exception flow analysis. Finally, we discuss the subject projects used.[1]

### A. An illustrative example of exception handling practices

In this subsection, we explain an illustrative example that handles, raises and propagates exceptions (see Figure 1). The example also illustrates the means of documenting exceptions.

*1) Handling possible exceptions:* In this example, a developer would like to implement a method named *A*. The method *A* requires to execute method *B*. The developer, by other means, has the knowledge that *B* can face two issues: 1) having an invalid path as input and 2) I/O faults. Therefore, instead of executing as expected, method *B* would possibly throw two types of exceptions: *InvalidPathException* and *IOException*, which correspond to the two issues, respectively. To deal with the two possible exceptions in method *B*, the developer needs to either handle the exception, i.e., determine the alternative actions when such exception happens, or propagate the exception such that a different method would manage the issue. In our example, the developer decides to handle *InvalidPathException* only and to propagate *IOException*.

---
[1]Source code, binaries, statistical tests and Tableau visualizations with raw data are available online at https://guipadua.github.io/scam2017.

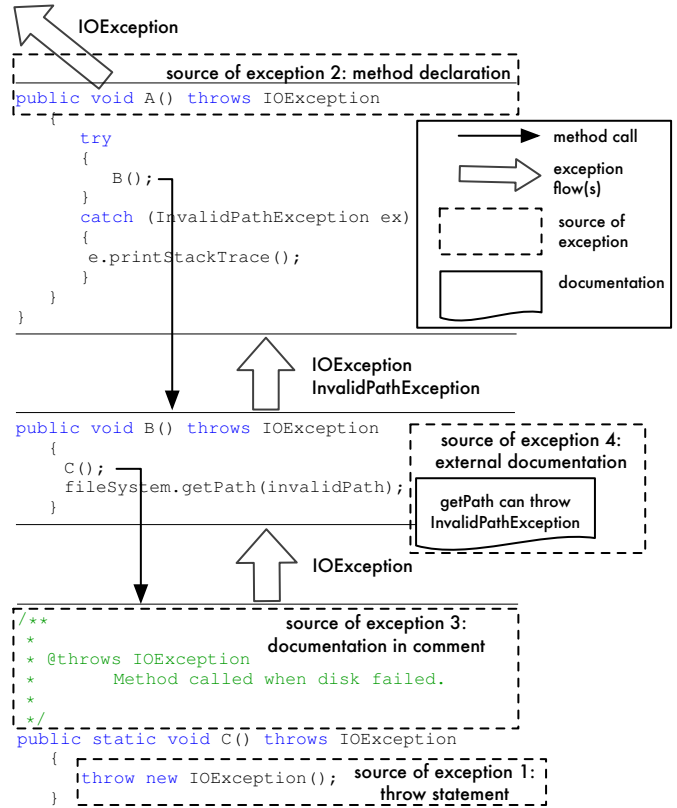

Fig. 1. An illustrative example.

*2) Raising and propagating exceptions:* As mentioned, the developer knows that *B* can have two issues, corresponding to two possible exceptions. These two exceptions can be either newly raised or propagated from another method that *B* calls (e.g., method *C* and method *getPath*). Developers can use a throw statement to raise an exception. In our example, developers newly raise *IOException* with the throw statement in the method *C*. Moreover, if the issue happens, such exception will propagate to all the methods that call *C*.

*3) Documenting exceptions:* From our example, a developer could consult the source code of *B* and *C* to identify the rise of *IOException* in the throw statement. However, the source code is not always available for a method. For example, the method *getPath* called in method *B* is declared by an external API. Developers would need to consult the documentation (such as JavaDoc) of the method to discover the propagation of the possible *InvalidPathException*.

There could be cases when an exception is thrown but still not available in the documentation. Therefore, in some programming languages (like Java), a possible exception can be part of the method declaration. For example, method *A*, *B*, and *C* declares that a possible exception can propagate in the throws block of the method declaration. However, some exceptions can remain undeclared (like *InvalidPathException*).

### B. Exception flow analysis

In the previous subsection, we presented an example of an exception handling scenario with its related flows. In this section, we present our methodology that automatically extracts possible exceptions and their flows in Java and C# projects. We build an automated tool using Eclipse JDT Core and .NET Compiler Platform ("Roslyn") to parse the Java and C# source code, respectively. As an overview, our analysis consists of three main steps. First, we identify the exception handling blocks (catch blocks). Second, we recover the flow of exceptions by constructing the call graph that is relevant to the identified catch blocks. Finally, by traversing the flow of exceptions, we identify the sources of possible exceptions.

As a building block, we obtain the abstract syntax tree (AST) from the source code. In this step, we include not only the source code but also the binary files of dependencies from the Java Virtual Machine (JVM) for Java or the .NET Global Assembly Cache (GAC) for C#. The dependencies of third party libraries used by the projects are also included. These dependencies enrich the analysis by providing *binding* information, which draws connections between the different parts of a program (i.e. any method call and its origin, either if part of an internal declaration or external dependency). Also, we enrich the AST by parsing the documentation of the dependencies mentioned above as another source of information.

*1) Identifying the handling of exceptions:* We collect all the exception handling scenarios through all the catch blocks available in the AST. At the catch block, we use the AST elements to identify the methods that are executed to handle each exception as handling actions. We also obtain the related try blocks, which provide a list of called methods. These methods are necessary since they might raise or propagate the exceptions that the catch block potentially handles. In our example, we identify the catch block in method *A*. The method *printStackTrace* is the handling action of the exception *InvalidPathException*. From this catch block, we obtain the try block in which we find the call to method *B*. Method *B* can potentially propagate *InvalidPathException* and *IOException*.

*2) Constructing call graph:* Exceptions are propagated in method calls. Therefore, we leverage call graphs to recover the flow of exceptions. To handle polymorphism without risking over-estimation, we only consider the possible exceptions of the method that is declared in the parent class, since they are more generic and often called within the derived methods. Based on the previous step, for each identified method we traverse its call graph in a depth-first manner to find its possible exceptions. In our example, we traverse the call graph of method *B* and find two possible exceptions: *IOException* from method *C* and *InvalidPathException* from method *getPath*. Hence, based on the examples' catch block, we know that the *InvalidPathException* is handled in method *A* while *IOException* is propagated without handling.

*3) Identifying sources of exceptions:* During the call graph traverse and based on the AST, we identify four sources of exceptions. They are: 1) The newly raised exception by the throw statement, 2) the declared exception in the throws of the method declaration (only for Java), 3) the documentation as comments in the source code (like JavaDoc comments), and 4) the external documentation. In our illustrative example, we can identify the newly raised *IOException* in a throw statement in method *C* (source 1), the declaration of the *IOException* in methods *A*, *B*, and *C* (source 2), and the JavaDoc documentation of method *C* for *IOException* (source 3). In addition, since we include the information from external libraries, our tool can also identify that the method *getPath* called in method *B* is a source of a possible *InvalidPathException* (source 4).

TABLE II
AN OVERVIEW OF THE SELECTED SUBJECT PROJECTS.

| | Project | Release Version | Type | # Try | # Catch | # Method (K) | KLOC |
|---|---|---|---|---|---|---|---|
| C# | Glimpse | 1.8.6 | App. | 56 | 57 | 1 | 31 |
| | Google API | v1.15.0 | Lib. | 22 | 30 | 16 | 628 |
| | OpenRA | release-20160508 | App. | 138 | 143 | 7 | 125 |
| | ShareX | v11.1.0 | App. | 334 | 341 | 7 | 177 |
| | SharpDevelop | 5.0.0 | App. | 940 | 1060 | 41 | 923 |
| | SignalR | 2.2.1 | Lib. | 94 | 105 | 2 | 38 |
| | Umbraco-CMS | release-7.5.0 | App. | 595 | 615 | 15 | 362 |
| Java | Apache ANT | rel/1.9.7 | App. | 934 | 1139 | 11 | 158 |
| | Eclipse JDT Core | I20160803-2000 | Lib. | 1,424 | 1655 | 25 | 383 |
| | Elasticsearch | v2.4.0 | App. | 385 | 408 | 12 | 108 |
| | Guava | v19.0 | Lib. | 263 | 317 | 10 | 79 |
| | Hadoop Common | rel/release-2.6.4 | Lib. | 975 | 1144 | 14 | 147 |
| | Hadoop HDFS | rel/release-2.6.4 | App. | 525 | 586 | 4 | 44 |
| | Hadoop MapReduce | rel/release-2.6.4 | App. | 293 | 367 | 6 | 57 |
| | Hadoop YARN | rel/release-2.6.4 | Lib. | 1,192 | 1529 | 29 | 257 |
| | Spring Framework | v4.3.2.RELEASE | Lib. | 1,940 | 2301 | 30 | 349 |
| | | | Total | 10,110 | 11,797 | 230 | 3,866 |

Some exceptions can be identified from multiple sources. For example, *IOException* is identified by three separate sources. We do not consider the multiple sources as different exceptions if the exceptions are associated with the same method call (e.g. method *B*). We label the separate sources of an exception as detailed information for each method call.

*C. Subject projects*

Table II depicts the studied subject projects. Our study considers Java and C# due to their popularity and prior research (see Section II). Moreover, we include C# because of its different approach compared to Java exception handling. To facilitate replication of our work, we chose open source projects that are available on GitHub.

We leverage GitHub filters on the number of contributors (i.e. projects with multiple contributors) and the number of stargazers (i.e. projects with more than ten stargazers), as they can achieve a good precision for selecting engineered software projects [29]. To narrow down the number of projects we also sorted the projects in descending order of the number of stargazers. Moreover, to potentially investigate the differences in exception handling practices and increase generalizability, we picked projects based on the filtering mentioned above. After reading the official description of the projects, we selected multiple applications and multiple libraries (i.e. project type), as well as multiple projects for different business domains (i.e. project purpose). From each project, we selected the most recent stable version of the source code at the moment of data collection for analysis.

## IV. QUANTITY OF EXCEPTIONS

In this section, we study the quantity of all possible exceptions that are in each try block.

### A. All and propagated possible exceptions

Ideally, developers should be aware of all possible exceptions to decide between handling or propagating them. To do that, developers need to navigate the call graph of a system that could extend to multiple ramifications. Hence, the more exceptions there are, the more challenging (i.e. exponential growth) it is for developers to comprehend and decide about exception handling. Besides that, missing possible exceptions can be a reason for the lack of a handler that should exist, which is considered one of the top causes of exception handlings bugs [8]. For those reasons, we study the quantity of total and propagated possible exceptions in each exception handling block.

As described in our methodology (see Section III), we collect all the methods called in each try block. For those methods, we can recover the possible exceptions. Afterward, we can measure the quantity of possible exception by counting the unique types of exceptions in each try block.

We find that there typically exist multiple possible exceptions in each try block (see Figure 2). The median number of distinct possible exception per try block is four and two, for C# and Java respectively. More than 48% (C#) and 38% (Java) of try blocks can throw in between two and five exceptions. Moreover, more than 36% (C#) and 24% (Java) of the try blocks have six or more possible exceptions. For example, an important method named *processCompiledUnits(int,boolean)* in Eclipse JDT Core in the *Compiler* class has a try block with 33 distinct possible exceptions. Developers should properly handle exceptions in such an important method, to ensure reliability.

Among all possible exceptions, there often exist possible exceptions that are not handled by any of the catch blocks that are associated with a try block [30]. These unhandled exceptions may increase the challenge of exception handling practice since they will be propagated and will need to be handled elsewhere. If the propagated exceptions remain uncaught across the whole system, there could be a risk of system failures [20], [7], [8]. Therefore, we identify the possible exceptions that are not handled by the catch block.

Figure 2 presents the possible propagated exceptions for each try statement. We find that there may exist a large number (up to 34) of possible exceptions that are unhandled in each try block. For example, a method *execute(List,int)* from class *org.apache.tools.ant.taskdefs.optional.junit.JUnitTask* in Apache ANT has 25 possible exceptions while the corresponding catch block only handles *IOException*.

> **Finding 1:** There often exist multiple possible exceptions in each try block, and, out of those, many are propagated.
> **Implications:** Current state-of-the-practice may not provide information to developers about all possible exceptions. Automated techniques may help developers be aware of all possible exceptions to make exception handling decisions.

### B. Potentially recoverable yet propagated exceptions

In the previous subsection, we find that a significant amount of the possible exceptions are propagated. However, not all exceptions are easy to recover or, more importantly, should even be recovered. For example, exceptions such as *ThreadDeath* in Java, and *OutOfMemoryException* in C# cannot feasibly be recovered. In fact, both Java and C# define the recoverability level of exceptions in their documentation [31], [32]. In particular, they suggest that developers should handle potentially recoverable exceptions while developers may not handle potentially unrecoverable ones. Hence, we first group all the propagated exceptions into either potentially recoverable or potentially unrecoverable, according to the specific guidance on Java or C# documentation. Then we count the number of propagated exceptions with potential recoverability.

We find that almost 8% (117) of C# and more than 19% (1,359) of Java try blocks have at least one potentially recoverable yet propagated exception. For example, a method named *rename* in Hadoop HDFS for file renaming features has a possible and potentially recoverable exception called *FileAlreadyExistsException*. This exception indicates the situation where a file is renamed to another existing file. However, this potentially recoverable exception is not handled by any catch block in that method.

> **Finding 2:** There exists a considerable amount of potentially recoverable exceptions that are propagated, even though they are recommended to be handled by Java and C#.
> **Implications:** Exception flow analysis can provide automated tooling support to alert developers about not handling potentially recoverable exceptions.

## V. DIVERSITY OF EXCEPTIONS

There can be a diverse set of exceptions being used across try blocks. Prior research discusses that the use of a high number of distinct exception types might represent a greater concern with exception handling [12]. Therefore, in this section, we study the diversity of exceptions in our subject projects.

We count the total number of distinct exception types in each project, and the amount of try blocks in which each type of exception appears. Table III shows the percentage of the exception types of each project that appear in different quantities of try blocks. Despite the large number (up to 97 in C# and 249 in Java) of distinct exception types, there exist a considerable amount of exception types that only appear in few try blocks. In fact, over half of the exception types in C#, and almost 1/3 of the exception types in Java only appear in one try block. Such results imply that although the high number of distinct exception types may be a burden to developers, the burden may not be as high since a considerable amount of the exception types would only affect a small portion of the code.

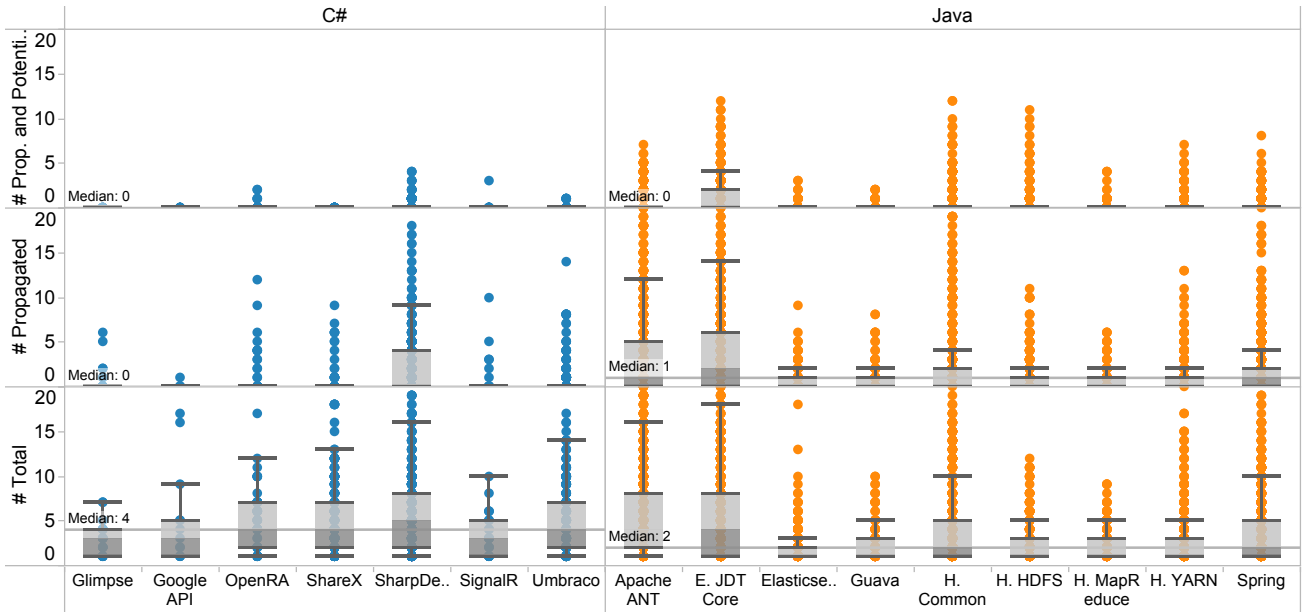

Fig. 2. Number of possible exceptions per try block for each project broken down by Propagated and Potentially Recoverable, Propagated, and Total.

TABLE III
TOTAL AMOUNT OF DISTINCT EXCEPTION TYPES AND THE PERCENTAGES OF DISTINCT EXCEPTION TYPES THAT APPEAR IN DIFFERENT QUANTITY OF TRY BLOCKS.

|     | Project | 1 | 2 | 3 | 4 | 5 | >5 | Total |
|---|---|---|---|---|---|---|---|---|
|     | Glimpse | 27.78% | 38.89% | 11.11% |  |  | 22.22% | 18 |
|     | Google API | 28.00% | 40.00% | 12.00% |  | 4.00% | 16.00% | 25 |
|     | OpenRA | 35.71% | 4.76% | 7.14% | 16.67% | 2.38% | 33.33% | 42 |
|     | ShareX | 13.04% | 8.70% | 6.52% | 2.17% | 6.52% | 63.04% | 46 |
| C#  | SharpDevelop | 19.59% | 8.25% | 6.19% | 4.12% | 2.06% | 59.79% | 97 |
|     | SignalR | 56.67% | 16.67% | 3.33% | 6.67% |  | 16.67% | 30 |
|     | Umbraco | 27.69% | 9.23% | 4.62% | 1.54% |  | 56.92% | 65 |
|     | **Total** | 55.88% | 25.00% | 13.97% | 9.56% | 5.15% | 47.79% | 214 |
|     | Apache ANT | 15.73% | 6.74% | 6.74% | 3.37% | 3.37% | 64.04% | 89 |
|     | E. JDT Core | 5.56% | 2.78% | 2.78% | 4.17% | 1.39% | 83.33% | 72 |
|     | Elasticsearch | 27.78% | 12.50% | 16.67% | 9.72% | 4.17% | 29.17% | 72 |
|     | Guava | 24.00% | 12.00% | 16.00% | 2.00% | 10.00% | 36.00% | 50 |
| Java| H. Common | 14.53% | 15.12% | 9.88% | 11.63% | 8.14% | 40.70% | 172 |
|     | H. HDFS | 27.50% | 13.75% | 16.25% | 7.50% | 1.25% | 33.75% | 80 |
|     | H. MapReduce | 21.74% | 8.70% | 4.35% | 8.70% | 2.17% | 54.35% | 46 |
|     | H. YARN | 17.53% | 4.12% | 11.34% | 6.19% | 3.09% | 57.73% | 97 |
|     | Spring | 22.09% | 12.05% | 7.63% | 10.04% | 4.82% | 43.37% | 249 |
|     | **Total** | 32.93% | 17.76% | 15.97% | 14.77% | 8.38% | 42.32% | 662 |
| ***Grand Total*** |  | 37.83% | 19.31% | 15.54% | 13.66% | 7.69% | 43.49% | 876 |

**Finding 3:** With a large amount of exceptions exist in each project, many possible exception types appear in only one try block.
**Implications:** Developers may not need to be aware of all exception types in a project by receiving automated suggestions of the exceptions that he/she needs to understand.

## VI. SOURCES OF EXCEPTIONS

The same exception may be traced back from different sources. In this section, we study the sources of exceptions per try block.

### A. Multiple sources of the same exception

The multiple sources of exceptions may increase the complexity of exception handling. Consequently, a developer would need to comprehend and investigate more methods in the source code to effectively handle exceptions. For example, developers may encounter a *FileNotFoundException* due to missing an input file or configuration file. However, developers may need different actions to handle such an exception since missing an input file may be caused by users' mistake while missing a configuration file is a critical issue of the software. Multiple sources of the same exception may also impact testers since they would need to properly test the exception behavior as well as the multiple possible paths of control flow.

We group each possible exception by the distinct methods that act as a source of exceptions. We only consider distinct methods since the same method may not need different ways to handle the exception while the exception propagated from various methods may need to be handled differently. In Table IV, we present the percentage of possible exceptions that are traced back from zero, one, two and more than two distinct methods. The first group is from zero methods, which means that these possible exceptions were traced back to explicit throw invocations, not method invocations.

Although most of the possible exceptions (above 76%) are traced back to a single method, we observe that more than 22% of the exceptions are traced back to multiple invoked methods. The try blocks with the highest number of methods can have from two to 17 among C# projects; while, for Java, it is between five and 34. For example, the Umbraco C# class called *TypeFinder* performs lazy accesses to all assemblies inside a single try block and therefore *System.ArgumentNullException* can be traced back from 14 different invoked methods. We also noticed that exceptions that are super classes of other exceptions (e.g., *IOException*) have a higher than average chances of being from multiple sources.

TABLE IV
PERCENTAGE OF DISTINCT EXCEPTIONS THAT ARE TRACED BACK TO ONE, TWO OR OVER TWO METHODS.

|  | Distinct methods | | | | Total |
|---|---|---|---|---|---|
|  | 0 | 1 | 2 | >2 |  |
| C# | 1.05% | 76.11% | 14.05% | 8.80% | 7,638 |
| Java | 0.61% | 77.18% | 13.00% | 9.20% | 28,854 |

**Finding 4:** Over 22% of the exceptions are traced from different methods.
**Implications:** Automated tools are needed to help developers understand the source of the exception if it is traced back to different methods.

### B. Sources of exception documentation

Prior studies revealed that lacking immediate documentation is one of the challenges of exception handling [14], [17], [18]. Prior studies observed a small number of documented exceptions. As shown in our illustrative example (see Section III-A), possible exceptions can be recovered from up to four different sources. They are: 1) the newly raised exception by the throw statement in the source code, 2) the declared exception in the throws of the method declaration (only for Java), 3) the documentation as comments in the source code (like JavaDoc comments) and 4) the external documentation. By recovering the sources of each possible exception using exception flow analysis, we may be able to provide the documentation of possible exceptions. For Java, we only recover documentation for unchecked exceptions since checked exceptions must be specified in method declarations.

We find that, from all the possible exceptions that we identify, 93% for Java and 71% for C# can be retrieved from the external documentation of dependencies. Figure 3 depicts the sets of possible exceptions per try block that were retrieved by our exception flow analysis. Our findings show that the challenge of having a low amount of documented exceptions can be well addressed by applying exception flow analysis with the information from external documents of libraries. We find that such rich documents are typically from the exceptions that are provided by the system libraries. Therefore, the high availability of such documentation can be expected to assist developers not only for our subject systems but also for the majority of Java and C# projects.

**Finding 5:** The libraries used by the systems can provide documentation to most of the possible exceptions.
**Implications:** Developers should leverage automated flow analysis to understand possible exceptions.

## VII. EXCEPTION HANDLING STRATEGIES AND ACTIONS

In this section, we study the strategy and actions in exception handling practices considering the exception flows of each handler.

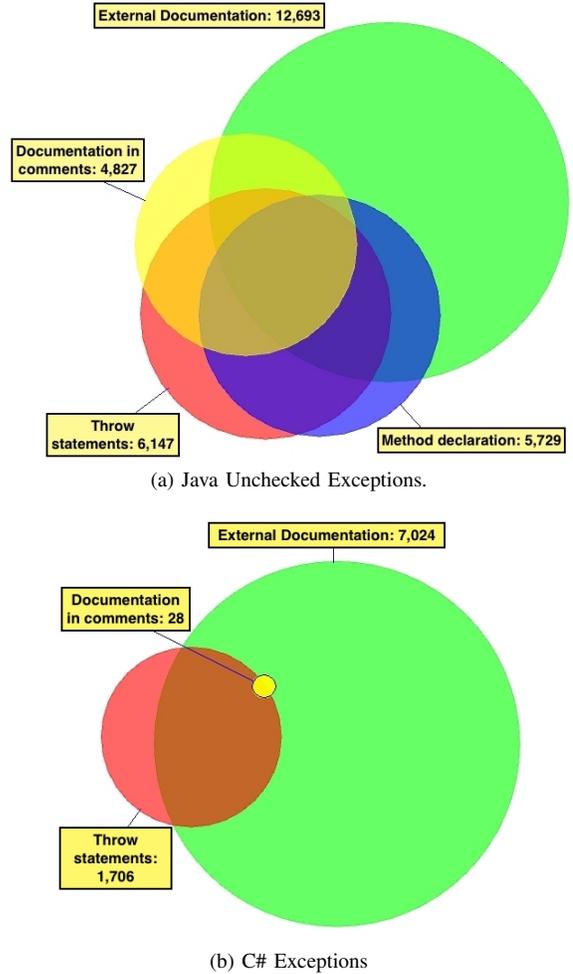

(a) Java Unchecked Exceptions.

(b) C# Exceptions

Fig. 3. Quantity of identified possible exceptions per source of information.

### A. Exception handling strategies

Exception handling strategy describes the manner in which an exception is handled. In particular, the relationship between the possible exception in a try block and the handler exception in the corresponding catch blocks. There exist in total two handling strategies:

- *Specific*, is the strategy when the type of a possible exception is exactly the same as the handler exception.
- *Subsumption*, is the strategy when the handler exception is a superclass of a possible exception.

Since there can be multiple possible exceptions, it can be overwhelming for a developer to handle each possible exception with a *Specific* strategy. On the other hand, the *Subsumption* strategy may introduce uncertainty to the caught exception. To study the handling strategies, we compare each possible exception with the handler exception in the corresponding catch block. Figure 4 depicts the quantity of distinct possible exception per try block that is handled according to each strategy.

The majority of the exceptions are handled with a subsumption strategy, while only a small portion of the exceptions are handled specifically. The results show that developers tend to over-catch exceptions. The extreme case of subsumption strategy is the "Catch Generic" exception handling anti-pattern [13], where developers simply use an exception type which can catch any exceptions in the software, e.g., *Exception* in Java. Such practice is heavily discussed in prior research [20], [17] and is considered to be harmful since developers cannot optimize the handling of the exception based on the exact type of the exception, but rather only know that there may exist some exceptions during run-time.

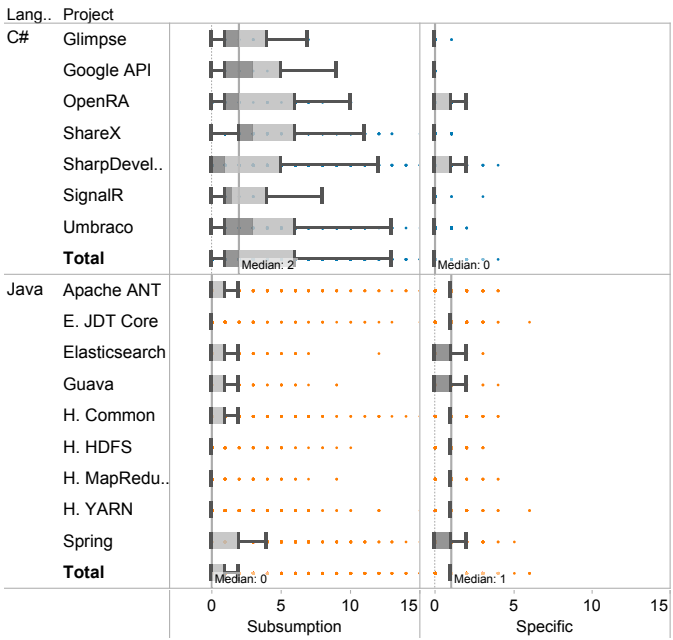

Fig. 4. Quantity of possible exceptions per try block per project by handling strategy.

> **Finding 6:** Only a small portion of the exceptions are handled with the *Specific* strategy.
> **Implications:** Developers should be guided to prioritize on handling exception with the *Specific* strategy, since developers cannot optimize the handling of an exception without knowing its exact type information.

### B. Exception handling actions

During the exception flow analysis (see Section III), we collect a set of method calls in each catch block to know how each exception is handled. Prior studies [3], [17], [14], [16], [28] propose a list of actions based on the combination of method calls in the catch block as Exception handling actions. Table V presents the list of actions that are defined in prior research and are used in this paper. To further understand how exceptions are handled, we study the exception handling actions in our subject projects.

Figure 5 presents the percentages of possible exceptions of each project that are handled using a particular type of action. We observe that Java and C# have differences in executing various actions when handling exceptions. To determine the differences, we perform Wilcoxon Rank Sum test [33] to compare the percentage of possible exceptions that are handled using each type of action in each C# and Java project. Hence, we examined if there exists statistically significant difference (i.e. p-value < 0.05) between Java and C#. A p-value < 0.05 means that the difference is likely not by chance. We choose Wilcoxon Rank Sum test since it does not have an assumption on the distribution of the data.

We find statistically significant difference between Java and C# for exception handing with actions "Throw Wrap", "Throw New", "Nested Try", "Continue" and "Todo". Among these actions, "Throw Wrap", "Throw New" and "Todo" may indicate that the exceptions are not effectively handled but rather propagated or ignored. All these three actions show up more in Java than C#. We consider the reason may be that Java compiler forces developers to explicitly manage checked exceptions while developers may not have the knowledge of how to handle them properly. To simply make the program compile, developers potentially take these actions. Further studies should investigate why such actions are chosen more in Java than C#.

> **Finding 7:** Java and C# have differences in leveraging various actions when handling exceptions.
> **Implications:** More in-depth analysis and user studies are needed to further understand the rationale of differences of Java and C# exception handling practices.

We also compare the actions that are taken when the exceptions are handled with either specific or subsumption strategy. We perform Wilcoxon Rank Sum test similar as when comparing Java and C#. This time, for each particular programming language and type of action, the test compared the percentage of specific handling in each project with subsumption handling in each project. Nonetheless, we observe only one action, i.e., *Log*, that is handled differently (statistically significant) with specific or subsumption strategy.

> **Finding 8:** Only one action, *Log* in Java, is taken differently when exceptions are handled with specific or subsumption strategy.
> **Implications:** Research and tooling support are needed to guide how to handle exceptions, especially with the specific strategy.

We would like to know if any particular actions are taken when handling some special possible exception. With such knowledge, we may be able to suggest actions automatically to developers handling exceptions. We gather a list of the ten most handled types of possible exceptions in Java and C#, respectively (see Table VI). We also obtain the percentage of possible exceptions that are handled using each action for each exception type. Similarly, we use Wilcoxon Rank Sum test to compare. For each particular programming language, action, and exception type, the test compared the percentage of the given type in each project with the combined value of all other types of exceptions in each project. We find with statistical significance that, for Java, all top exceptions have at least one action that is taken differently from the rest of the exceptions, and, for C#, two top exceptions has such difference.

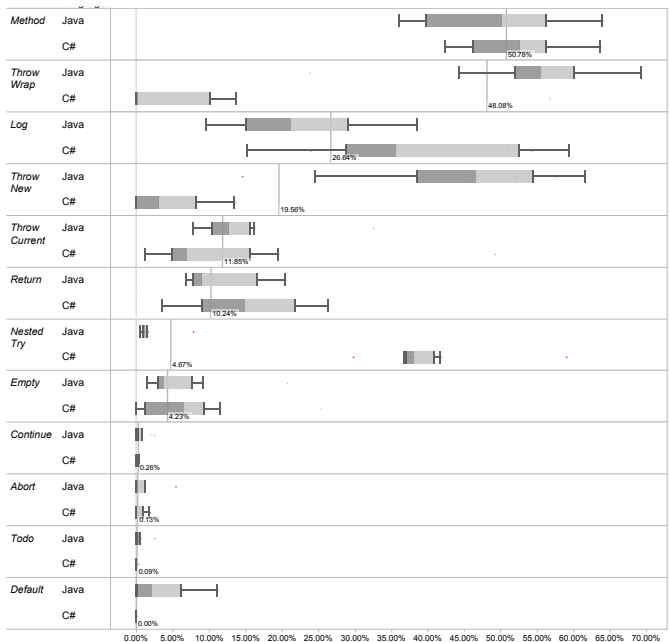

Fig. 5. Percentage of possible exceptions that are handled using each type of action.

**Finding 9:** All top 10 Java and 2 out of top 10 C# exceptions have at least one action that is taken statistically significantly differently from the rest exceptions.
**Implications:** Developers may consider leveraging automated suggestions of exception handling actions.

TABLE V
LIST OF THE DETECTED ACTIONS.

| Action | Short Description |
|---|---|
| Abort | The handler contains an abort statement [3]. |
| Continue | The handler contains a continue statement [14]. |
| Default | The handler contains the IDE suggested method (Java only). |
| Empty | The handler is empty [3], [17], [14], [16]. |
| Log | The handler displays or logs some information [17], [14], [16]. |
| Method | The handler contains a method invocation different than the other actions listed in this table. [14], [17]. |
| Nested Try | The handler contains a new try statement [28]. |
| Return | The handler contains a return statement [17], [14]. |
| Throw Current | The handler contains a throw statement without a new exception instantiation [17], [14], [16]. |
| Throw New | The handler contains a throw statement with a new exception instantiation [17], [14], [16]. |
| Throw Wrap | The handler contains a throw statement using the original exception or its associated information [16]. |
| Todo | The handler contains TODO or FIXME comments [3]. |

## VIII. THREATS TO VALIDITY

In this section, we discuss the threats to validity of our findings.
**External validity.** Our study is based on a set of open source Java and C# projects from GitHub. Our findings may not generalize to other projects, languages or commercial systems. Replicating our study on other subjects may address this threat and further understand the state-of-the-practice of exception handling.
**Internal validity.** We aim to include all possible sources of information in our automated exception flow analysis. However, our analysis may still miss possible exceptions, if there is a lack of documentation or the source code is not compilable. Also, the documentation of the exception may be incorrect or outdated. In our analysis, we trust the content of documentation. Therefore, we cannot claim that our analysis fully recovers all possible exceptions nor that the recovered information is impeccable. Further studies may perform deeper analysis on the quality of exception handling documentation to address this threat.
**Construct validity.** Our study may not cover all possible handling actions. We selected actions based on the previous research in the subject [3], [17], [14], [16], [28]. Some actions are not included in our study if they are either 1) require heuristic to detect or 2) are not well explained in details in related work.

TABLE VI
LIST OF TOP 10 COMMON EXCEPTION TYPES IN THE STUDIED PROJECTS.

| C# | Java |
|---|---|
| System.ArgumentNullException | java.io.IOException† |
| System.ArgumentException | java.lang.IllegalArgumentException† |
| System.NotSupportedException† | java.lang.NullPointerException† |
| System.ArgumentOutOfRangeException | java.lang.IndexOutOfBoundsException† |
| System.InvalidOperationException | java.lang.SecurityException† |
| System.FormatException | java.lang.IllegalStateException† |
| System.IO.IOException | java.lang.ExceptionInInitializerError† |
| System.IO.PathTooLongException† | java.lang.ArrayStoreException† |
| System.Security.SecurityException | java.lang.IllegalAccessException† |
| System.ObjectDisposedException | java.lang.ClassNotFoundException† |

†The exception has at least one actions that taken statistically significantly differently from the rest exceptions.

Our possible exception identification approach is based on a call graph approximation from static code analysis. We may still miss possible exceptions due to under-estimation for polymorphism or unresolved method overload. Although such approximation may impact our findings, our choice of under-estimation would not significantly alter the existence of observed challenges of exception handling, i.e., the challenge may appear even worse without the under-estimation. Nevertheless, to complement our study, dynamic analysis on the exception flow may be carried out to understand the system exceptions during run-time.

## IX. CONCLUSION

Exception handling is an important feature in modern programming languages. However, prior studies unveil the suboptimal usage of exception handling features in practice. In this paper, we revisit practice of exception handling in 16 open source software in Java and C#. Although we confirm that there exist suboptimal manners of exception handling, more importantly, we highlight the opportunities of performing source code analysis to recover exception flows to help practitioners tackle various complex issues of handling exceptions. In particular, the contributions of our paper are:

1) We design an automated tool that recovers exception flows from both Java and C#.
2) We present empirical evidence to illustrate the challenges and complexity of exception handling in open source systems.
3) Our exception flow analysis, as an automated tool, can already provide valuable information to assist developers better understand and make exception handling decisions.

This paper highlights the opportunities and urgency of providing automated tooling to help developers make exception handling decisions during the development of quality and reliable software systems.